\documentclass[12pt,aps,prd,amsfonts,preprintnumbers,superscriptaddress,showpacs,nofootinbib]{revtex4}
\usepackage{epsfig,amssymb,amsfonts,bm}

\newcommand{\vep}{{\bm p}}
\newcommand{\vek}{{\bm k}}
\newcommand{\veq}{{\bm q}}

\newcommand{\be}{\begin{equation}}
\newcommand{\ee}{\end{equation}}
\newcommand{\bea}{\begin{eqnarray}}
\newcommand{\eea}{\end{eqnarray}}
\newcommand{\beas}{\begin{eqnarray*}}
\newcommand{\eeas}{\end{eqnarray*}}
\newcommand{\ds}{\displaystyle}
\newcommand{\0}{^{\rm{ph}}}
\newcommand{\oo}[1]{(#1^{\rm ph})^2}

\DeclareMathSymbol{\varGamma}{\mathord}{letters}{"00}
\begin{document}

\title{Binding energy of the $X(3872)$  at unphysical pion masses}

\author{V. Baru}
\affiliation{Institut f\"ur Theoretische Physik II, Ruhr-Universit\"at Bochum,
D-44780 Bochum, Germany}
\affiliation{Institute for Theoretical and Experimental Physics, B. Cheremushkinskaya 25, 117218 Moscow, Russia}

\author{E. Epelbaum}
\affiliation{Institut f\"ur Theoretische Physik II, Ruhr-Universit\"at Bochum,
D-44780 Bochum, Germany}

\author{A.A. Filin}
\affiliation{Institut f\"ur Theoretische Physik II, Ruhr-Universit\"at Bochum,
D-44780 Bochum, Germany}

\author{J. Gegelia}
\affiliation{Forschungszentrum J\"ulich, Institute for Advanced Simulation, Institut f\"ur Kernphysik and
J\"ulich Center for Hadron Physics, D-52425 J\"ulich, Germany}
\affiliation{Tbilisi State University, 0186 Tbilisi, Georgia}

\author{A.V. Nefediev}
\affiliation{Institute for Theoretical and Experimental Physics, B. Cheremushkinskaya 25, 117218 Moscow, Russia}
\affiliation{National Research Nuclear University MEPhI, 115409, Moscow, Russia}
\affiliation{Moscow Institute of Physics and Technology, 141700, Dolgoprudny, Moscow Region, Russia}

%\date{\today}

\begin{abstract}
Chiral extrapolation of the $X(3872)$ binding energy is investigated using the modified Weinberg formulation of chiral
effective field theory for the $D \bar{D}^*$ scattering. Given its explicit renormalisability, this approach is
particularly useful to explore the interplay of the long- and short-range $D \bar{D}^*$ forces in the $X(3872)$ from studying the
light-quark (pion) mass dependence of its binding energy. In
particular, the parameter-free leading-order calculation shows that the $X$-pole disappears for unphysically large pion
masses. On the other hand, without contradicting the naive dimensional analysis,
the higher-order pion-mass-dependent contact interaction can change the slope of the binding energy at the
physical point yielding the
opposite scenario of a stronger bound $X$ at pion masses larger than its physical value. An important role of the pion
dynamics and of the 3-body $D\bar{D}\pi$ effects for chiral extrapolations of the $X$-pole is emphasised. The results of
the
present study should be of practical value for the lattice simulations since they provide a non-trivial connection
between lattice points at unphysical pion masses and
the physical world.
\end{abstract}


\maketitle

\section{Introduction}

After more than a decade from the discovery by the Belle Collaboration of the $X(3872)$ charmonium-like state
\cite{Choi:2003ue} its
nature still remains an open question --- see Ref.~\cite{QWGreview} for a review. According to the Particle Data Group,
this state has the mass
$M_X$=(3871.68~$\pm$~0.17)~MeV \cite{PDG} and thus resides very close to the
neutral $D \bar{D}^* $ threshold with
\be
E_B=M_{D^0}+M_{\bar{D}^{*0}}-M_X= (0.12 \pm 0.26 )\ {\rm MeV}.
\ee
It is therefore natural to assume that its wave function has a large molecular admixture --- see a vast literature
on hadronic molecules, for example,
Refs.~\cite{Weinberg:1962hj,Voloshin:1976ap,DeRujula:1976qd,tornqvist,Voloshin:2003nt,Tornqvist:2004qy,swanson,wong,
Baru:2003qq,Braaten:2007dw,Hanhart:2007yq,Kalashnikova:2009gt} to mention some, also in the context of the
$X(3872)$.

The $1^{++}$ quantum numbers of the $X$ determined recently by the LHCb Collaboration~\cite{Aaij:2013zoa,Aaij:2015eva} are
consistent with its interpretation as an $S$-wave $D^0 \bar{D}^{* 0}$ bound state\footnote{A proper $C$-parity
eigenstate is always meant by this (and similar) shorthand notation.} --- see, for example,
Refs.~\cite{Voloshin:2003nt,Braaten:2007dw,Kalashnikova:2009gt}.
The small binding energy relative to the $D^0 \bar{D}^{* 0}$ threshold allows for
an effective field theory (EFT) formulation of the problem in analogy to the
deuteron. The pionless EFT based on pure contact $D\bar{D}^*$ interactions was first applied to the $X(3872)$ in
Ref.~\cite{AlFiky:2005jd} while implications of the heavy quark and heavy flavour symmetries were utilised in
Refs.~\cite{Guo:2013sya,Nieves:2012tt} to predict partner states of the $X(3872)$.
However in presence of other relevant dynamical scales such a treatment is expected to be
valid only in a very narrow energy region around the threshold. In particular, the 3-body neutral channel
$D^0 \bar{D}^0\pi^0$ opens at approximately 7 MeV below the $D^0\bar{D}^{*0}$ threshold while the charged 3-body
thresholds
$D^\pm D^\mp \pi^0$ and $D^\pm\bar{D}^0\pi^\mp$ reside about 2~MeV above it. In addition, the
charged two-body threshold $D^\pm \bar{D}^{*\mp}$ is located around 8~MeV above the neutral one.   The mass difference between the charged and 
neutral $D\bar D^*$ thresholds  was  shown 
  in Ref.~\cite{Gamermann:2009fv}    to play a  crucial role for understanding of isospin violation in the decays  of the $X$  into
  $\pi^+\pi^- J/\psi$    ~\cite{Belle2pi} and $\pi^+\pi^-\pi^0 J/\psi$ \cite{delAmoSanchez:2010jr}, for which   approximately equal branching 
  fractions were observed.
To incorporate the long-range pion physics the so-called X-EFT approach was developed in Ref.~\cite{Fleming:2007rp}
based on the assumption that pions can be treated perturbatively. Recently this framework was extended to include
higher-order corrections and then used to predict
the pion-mass dependence of the $X$-pole \cite{Jansen:2013cba} and the finite volume corrections to the $X$ binding
energy~\cite{Jansen:2015lha}. On the other hand, perturbative treatment of pions has a smaller range of validity
compared to nonperturbative approaches and it has to be used with caution---for example, the perturbative framework is
known to be not applicable in the deuteron channel \cite{Fleming:1999ee} which demonstrates certain similarities with
the $X(3872)$.

The frameworks with nonperturbative pions were employed in many phenomenological studies --- see, for example,
Refs.~\cite{molecule3,ThCl,Liu:2008fh} to mention some --- all of them, however, include one-pion-exchange (OPE) in the
static limit, that is under the assumption that the $D$-mesons are infinitely heavy.
Meanwhile, a close proximity of the $D^\pm D^\mp \pi^0$, $D^\pm\bar{D}^0 \pi^\mp$, and
$D^0 \bar{D}^0 \pi^0$ thresholds to the $X(3872)$ pole suggests that 3-body scales can play an
important role in this state, so that neglecting the 3-body dynamics one distorts the analytical
structure of the amplitude in the kinematical region of interest. It has to be noticed, however, that the proper
inclusion of the 3-body dynamics requires special care. For example, it is shown in Ref.~\cite{deeply} that
the 3-body unitary cuts play very important role in the $D_{\alpha}\bar{D}_{\beta}$ system, if the $D_{\beta}$
width is dominated by the $S$-wave $D_{\beta}\to D_{\alpha}\pi$ decay. In particular, it is demonstrated that
if the $D_\beta\to D_\alpha \pi$ coupling is sufficiently strong to produce a bound state \cite{close1,Close:2010wq}, it
is, at the same time, necessarily sufficiently strong to provide the state with such a large width that it
becomes unobservable. In turn, in the case of $P$-wave vertices, the system at
hand demonstrates additional difficulties since the selfenergy loops diverge, so that the system requires a proper
treatment to avoid false conclusions --- see an example of such conclusions in Ref.~\cite{Wang:2013kva} and its detailed
discussion in Ref.~\cite{Baru:2015nea}.  In particular,    contrary to  the  claims of Ref.~\cite{Wang:2013kva},
 it is shown  in Ref.~\cite{Baru:2015nea} that  the one-pion exchange (OPE) potential in the  $\bar D D^*$ system
  is well defined in the sense of an effective field theory only in connection with a contact operator.

In Refs.~\cite{Baru:2011rs,Baru:2013rta}, the properties of the $X(3872)$ molecular state were studied in a heavy-meson
EFT framework with nonperturbative pions including all relevant 3-body scales. It was understood that the
dynamical treatment of pions had a big impact on the $X$ line shape and, in particular, on the partial decay width
$X\to D^0\bar{D}^0\pi^0$. Furthermore, it was shown in Ref.~\cite{Baru:2011rs} that the static OPE approximation was not
adequate to analyse the role of the long-range pion dynamics in the $X(3872)$, since it corresponded
to an uncontrolled modification of the proper dynamical scales related to the $D\bar{D}\pi$ cuts and to the neglect of
the imaginary part of the $D\bar{D}^*$ potential. Meanwhile, the role of nonperturbative effects for these observables
appeared to be quite moderate, as follows from the agreement between the results of the nonperturbative calculations
\cite{Baru:2011rs} and those in the X-EFT \cite{Voloshin:2003nt,Fleming:2007rp}.

In Ref.~\cite{Baru:2013rta}, the nonperturbative framework developed in Ref.~\cite{Baru:2011rs} was generalised to
study the dependence of the $X$ binding energy on the light-quark mass or, equivalently, on the pion mass.
The use of non-perturbative one-pion exchange  for chiral extrapolations allows one to extend the region of
applicability of the approach to larger pion masses
which is important for analysing the results of lattice QCD calculations.

In this work we address another important issue which is related to nonperturbative
renormalisation of the 3-body Lippmann-Schwinger or Faddeev-type equations to describe the
interaction between heavy mesons in the $X$. The standard nonrelativistic approach to heavy mesons
leads to coupled-channel integral equations for the scattering amplitudes which, at leading order in the EFT
expansion, are linearly divergent. As a consequence, iterations of the truncated potential within the dynamical equation
generate an infinite series of ultraviolet (UV) divergent higher-order contributions to the amplitude which cannot be
absorbed into a finite number of counter terms (contact interactions) included in the potential. In other words, the
coefficients in front of the logarithmic and power-law divergences appearing in the iterations of the equation
involve powers of external momenta which can only be removed if an infinite number of higher-order (derivative)
contact interactions is included. The standard way to deal with this problem is to employ a
finite UV cut-off of the order of a natural hard scale in the problem which would suppress the unwanted higher-order
contributions, as advocated in Ref.~\cite{Lepage:1997cs}. This strategy was followed, in particular, in
Refs.~\cite{Nieves:2012tt,Valderrama:2012jv,Guo:2013sya,Baru:2011rs,Baru:2013rta}. Exactly
the same problem with renormalisation emerges also in the context of nuclear chiral EFT --- see, for example,
Refs.~\cite{Nogga:2005hy,Epelbaum:2006pt} and references therein. In particular, a finite cut-off was
employed for the construction of the NN potential and the few-body nuclear forces within chiral EFT --- see
Ref.~\cite{Epelbaum:2008ga} for a review. This procedure induces cut-off artefacts which might turn to a nontrivial
issue, in particular, for chiral extrapolations since it might be difficult to control the pion mass dependence of
short-range interactions in a systematic way. Note also that in Ref.~\cite{Baru:2013rta} the $m_{\pi}$-dependence of the
contact interaction was promoted to the leading order to maintain the renormalisability of the scattering amplitude at
unphysical pion masses.

Recently a novel, renormalisable (in the EFT sense) approach to nucleon-nucleon scattering with nonperturbative
pions was proposed in Ref.~\cite{Epelbaum:2012ua}. Starting from the Lorentz invariant form of the effective
Lagrangian, the authors of Ref.~\cite{Epelbaum:2012ua} derived a 3-dimensional dynamical equation which complies with
the relativistic elastic unitarity and which is renormalisable at leading order of EFT. Indeed, in the suggested
approach, all logarithmically divergent contributions generated by iterations of the potential can be fully absorbed
into the redefinition of the leading-order contact terms. Then higher-order contributions are subject to a
perturbative treatment in this approach. It should be stressed that the central point of the approach is
noncommutativity of the nonrelativistic expansion and the renormalisation procedure while after
renormalisation relativistic effects as such provide only minor impact on the low-energy observables, as it should be in
EFT. Apart from its transparency with regard to renormalisation, this approach allows one to remove finite cut-off
artefacts and it is very well suited for carrying out chiral extrapolations and studying correlations between the
effective
range parameters induced by the analytic structure of the long-range forces --- see
Refs.~\cite{Epelbaum:2013ij,Baru:2015ira} for the corresponding results in the NN sector. Given the
same UV behaviour of the dynamical equations for NN and heavy meson-antimeson scattering, we demonstrate that the method
of Ref.~\cite{Epelbaum:2012ua} can be used to reformulate the nonrelativistic 3-body approach of
Refs.~\cite{Baru:2011rs,Baru:2013rta} in terms of renormalisable integral equations. We apply the resulting
theoretical framework to study the quark mass dependence of the $X(3872)$ binding energy. In contrast to the finite
cut-off formulation, the $m_{\pi}$-dependence of the binding energy is predicted at leading order in
a renormalisable approach.

The paper is organised as follows. In Sec.~\ref{Sec2} we give a brief introduction to the method suggested in
Ref.~\cite{Epelbaum:2012ua}. In Sec.~\ref{equations} the problem of the $D\bar{D}^*$ interaction is formulated in
a closed selfconsistent form which makes it possible to appeal to the approach discussed in Sec.~\ref{Sec2}. In
Sec.~\ref{results} we present and discuss the results of our calculations. We summarise our findings in
Sec.~\ref{summary}. All necessary technical details are collected in Appendix~\ref{app:renorm}.

\section{Non-relativistic (Lippmann-Schwinger) equations versus equations with relativistic unitarity}\label{Sec2}

A proper nonrelativistic expansion of low-energy physical quantities can be done by calculating these quantities in a
Lorenz-invariant theory and expanding the final result in the powers of the velocity $v$ (we work in the natural system
of units setting the speed of light $c=1$), see  e.g.  Ref.~\cite{Bernard:1992qa}  where this issue is discussed in the one-nucleon sector. 
On the other hand, one can perform the nonrelativistic
expansion at the level of the Lagrangian of the theory. However, this expansion does not commute with the loop
integration. This can be exemplified
by a simple calculation adapted from Ref.~\cite{Epelbaum:2012cv}. Consider a scalar two-point loop function which is
logarithmically divergent and therefore should be regularised. With the simplest regularisation prescription given
by a sharp cut-off in the 3-dimensional momentum it reads
\be
I=\frac{4i}{(2\pi)^4}\int\frac{d^4 k \;\theta(\Lambda-|\vek|)}{\left[k^2-m^2+i0\right]\left[(P-k)^2-m^2+i0\right]},
\label{Iex}
\ee
where ${P=(2 \sqrt{m^2+\vep^2},{\bm 0})}$. The integral can be evaluated analytically for ${\Lambda>|\vep|}$
with the result
\be
I=-\frac{i |\vep|}{2\pi\sqrt{m^2+\vep^2}}
+\frac{|\vep|}{\pi^2\sqrt{m^2+\vep^2}}\ln\frac{\Lambda\sqrt{m^2+\vep^2}
+|\vep|\sqrt{\Lambda^2+m^2}}{m\sqrt{\Lambda^2-\vep^2}}
-\frac{1}{\pi^2}\ln\frac{\Lambda+\sqrt{\Lambda^2+m^2}}{m}.
\label{Ires}
\ee

A nonrelativistic expansion of the integrand in Eq.~(\ref{Iex}) implies the strong inequality
$|\vep|\ll\Lambda\ll m$ that is equivalent to the $1/m$ expansion made prior to the $1/\Lambda$ expansion in the exact
result (\ref{Ires}), that yields
\be
I=-\frac{i|\vep|}{2\pi m}-\frac{\Lambda}{\pi^2 m}+\ldots,
\label{sexp}
\ee
where the ellipsis denotes suppressed terms. Divergence in expression (\ref{sexp}) is linear, that is, it is stronger
than that in the original integral (\ref{Iex}) which is a consequence of the nonrelativistic expansion of the integrand.

On the contrary, keeping the integrand relativistic and performing the nonrelativistic expansion after integration is
equivalent to imposing a different, and more natural, strong inequality $|\vep|\ll m\ll\Lambda$ and, therefore, the
$1/\Lambda$ expansion is to be
performed in Eq.~(\ref{Ires}) before the $1/m$ expansion. This leads to a different result for the real part of the
integral,
\be
I=-\frac{i|\vep|}{2\pi m}-\frac{1}{\pi^2}\ln\frac{2\Lambda}{m}+\ldots,
\label{fexp}
\ee
which reveals the logarithmic divergence, in agreement with the UV behaviour of the original integral.

Thus nonrelativistic expansion of the integrand changes its ultraviolet behaviour and the final result differs from
the relativistic expansion of the exact expression for the integral. This difference is caused by noncommutativity of
the nonrelativistic
expansion and the loop integration. Because of this noncommutativity, in order to reproduce the results of the
Lorentz-invariant theory, one needs to add compensating terms to the nonrelativistic effective Lagrangian.
Therefore, more singular behaviour of the nonrelativistic equation leads to perturbative nonrenormalisability already
for the leading order (LO)
potential. In particular, iterations in the Lippmann-Schwinger equation generate power-law divergences with
coefficients of a progressively increasing power of the momentum/energy. As the LO potential does not contain
momentum-dependent
contact interactions, one cannot get rid of these divergences by absorbing them into redefinition of the parameters of
the LO
potential. Adding any finite number of momentum/energy dependent terms does not resolve the issue.
While this is not a problem when calculating a finite number of diagrams, it is rather disturbing when solving
integral equations. Except for some trivial
cases, it is not possible to take into account contributions of an infinite number of compensating
terms required for ``correcting'' an infinite number of iterations. One is, therefore, forced either to keep the
ultraviolet cut-off
finite ($\Lambda\sim m$) or to resort back to the original Lorentz-invariant formulation of the theory, although the
effect of relativistic corrections at low energies is, of course, small after renormalisation.

On the other hand, iterations of the Lippmann-Schwinger equation without nonrelativistic expansion generate only
logarithmic
divergences. This guarantees a perturbative renormalisability of the theory at LO, that is, all divergences can be
removed by renormalising the coupling constant of the LO contact interaction.

In Ref.~\cite{Baru:2013rta}, the nonrelativistic Lippmann-Schwinger equation was solved that corresponds to the
nonrelativistic expansion of the integrand as was explained above. In the present paper we deal with the relativised
Lippmann-Schwinger equation and, therefore, the LO amplitude is obtained by solving a renormalisable integral equation.
Analogously to the nucleon-nucleon scattering in the modified Weinberg approach of Ref.~\cite{Epelbaum:2012ua},
the relativised integral equation for the $D\bar{D}^*$ system has a milder ultraviolet
behaviour if compared to the nonrelativistic Lippmann-Schwinger equation.
It has to be noticed, however, that the integral equation becomes nonrenormalisable if higher-order corrections to the
leading-order potential are
also treated nonperturbatively. In particular, by iterating higher-order
contributions in the potential one generates divergences with such structures of momentum- and/or energy-dependent
coefficients which are not present in the iterated potential, that is these divergent contributions cannot be absorbed
into the redefinition of the contact terms
included in the potential at the given order. On the other hand, renormalisability is retained by
treating corrections perturbatively. In particular, if we denote the LO amplitude as $T_0$ and the NLO corrections to it
as $T_1$, that is we have the following perturbative expansion of the full amplitude:
\be
T=T_0+\varepsilon\, T_1+O(\varepsilon^2 ),
\label{Tampl}
\ee
then the inverse amplitude takes the form
\be
T^{-1}=T_0^{-1}(T_0-\varepsilon\, T_1)T_0^{-1}+O(\varepsilon^2 T_0^{-1}),
\label{pertT}
\ee
where $\varepsilon$ stands either for the expansion parameter of chiral EFT, $\varepsilon\sim \{m_{\pi}/\Lambda_{\chi},
p/\Lambda_{\chi}\}$ with $\Lambda_{\chi}$ being the
chiral symmetry breaking scale, or it corresponds to the
expansion around the physical pion mass $m_\pi\0$, that is $\varepsilon\sim (m_\pi^2-{m_\pi\0}^2)/{m_\pi\0}^2$.
In what follows, while we stick to the leading-order chiral potential,
we investigate the pion-mass dependence of the $X(3872)$ binding energy including corrections at the NLO which appear
as one goes away from the physical point.

Note that the expression of Eq.~(\ref{pertT}) gives an explicitly unitary amplitude, however it also includes
selectively resumed higher-order contributions which do not affect renormalisability of the scattering amplitude.
A bound state corresponds to the zero of the inverse amplitude (\ref{pertT}).
Finally, we use the superscript ``ph'' to label quantities taken at the physical point, that is for $m_\pi=m_\pi\0$.

\section{System of coupled-channel integral equations for the $D\bar{D}^*$ problem}
\label{equations}

In this Section we outline briefly our theoretical formulation of the problem. We follow the lines of
Ref.~\cite{Baru:2013rta} adapting the approach according to Ref.~\cite{Epelbaum:2012ua}.

The lowest-order $D^*D\pi$ interaction Lagrangian is taken in the form \cite{Fleming:2007rp}
\begin{eqnarray}
{\cal L}=\frac{g_c}{{2}f_\pi}\left({\bm D^*}^\dagger \cdot {\bm\nabla}{\pi^a}\tau^a D
+D^\dagger\tau^a{\bm\nabla}{\pi^a} \cdot {\bm D^*}\right).
\label{lag}
\end{eqnarray}
The dimensionless coupling constant $g_c$ is related to the $D^{*0}\to D^0\pi^0$ decay width as
\be
\varGamma(D^{*0}\to D^0\pi^0)=\frac{g_c^2m_0 q^3}{24\pi f_\pi^2m_{*0}},
\label{D*width2}
\ee
where $q=\lambda^{1/2}(m_{*0}^2,m_0^2,m_{\pi^0}^2)/(2m_*)$ is the center-of-mass 3-momentum of the outgoing particles
and $\lambda(x,y,z)$ is the standard triangle function --- see the definition in Eq.~(\ref{lambda}). Here and in what
follows, $m_*$, $m$, and $m_\pi$ denote the masses of the $D^*$ meson, $D$ meson, and pion, respectively. Charged and
neutral states are distinguished by an additional index, for example $m_{*c}$ versus $m_{*0}$.

\begin{figure}[t]
\centerline{\epsfig{file=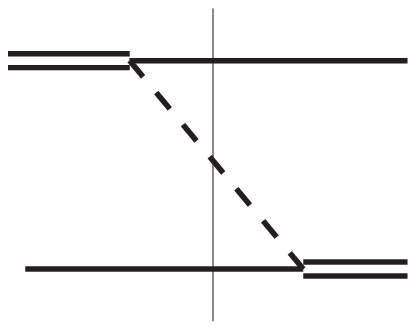, width=0.2\textwidth}\hspace*{0.1\textwidth}\epsfig{file=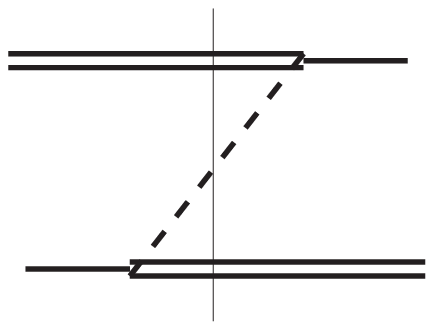, width=0.2\textwidth}}
\caption{Diagrams in time-ordered perturbation theory corresponding to the (inversed) 3-body propagators
$D^{(1)}_{3ij}$ (left plot) and $D^{(2)}_{3ij}$ (right plot) ($i,j=0,c$). The double and single solid lines refer to
the $D^*$ and $D$, respectively, while the dashed lines refer to pions. The thin vertical line pinpoints the
intermediate state.}\label{fig:D1D2}
\end{figure}

The $D\bar{D}^*$ potential at  LO in Chiral Effective Field Theory (ChEFT) consists of the
OPE and the $S$-wave derivativeless contact interaction $C_0$,
\be
V^{nn'}_{ij}(\vep,\vep')={(\vep+\vep')^n(\vep+\vep')^{n'}} F_{ij}(\vep,\vep')+ C_0\, \delta^{nn'},
\label{Vmn}
\ee
where indices $n$ and $n'$ are contracted with the corresponding indices of the $D^*$ polarisation vectors. Here
\be
F_{ij}(\vep,\vep')=-\frac{g_c^2}{(4\pi f_\pi)^2}\left(\frac{1}{D_{3ij}^{(1)}(\vep,\vep')}+\frac{1}{D_{3ij}^{(2)}(\vep
,\vep')}\right),
\label{Fij}
\ee
and $D^{(1)}_{3ij}$ and $D^{(2)}_{3ij}$ ($i,j=0,c$) stand for the $D\bar{D}\pi$ and $D^*\bar{D}^*\pi$ propagators
written in the framework of the Time-Ordered Perturbation Theory (TOPT) --- see Fig.~\ref{fig:D1D2},
\be
D_{3ij}^{(1)}(\vep,\vep')=
\left\{
\begin{array}{lcl} 
\ds{E_{\pi^0}(\vep+\vep')}\Bigl(E_{D_i}(p)+E_{D_i}(p')+E_{\pi^0}(\vep+\vep')
-M\Bigr),&& i=j,\\[4mm]
\ds{E_{\pi^c}(\vep+\vep')}\Bigl(E_{D_i}(p)+E_{D_j}(p')+E_{\pi^c}(\vep+\vep')
-M\Bigr),&& i\ne j,
\end{array}
\right.
\label{D1}
\ee
\be
D_{3ij}^{(2)}(\vep,\vep')=
\left\{
\begin{array}{lcl} 
\ds {E_{\pi^0}(\vep+\vep')} \Bigl(E_{D^*_i}(p)+E_{D^*_i}(p')+E_{\pi^0}
(\vep+\vep')-M\Bigr),&& i=j,\\[4mm]
\ds {E_{\pi^c}(\vep+\vep')} \Bigl(E_{D^*_i}(p)+E_{D^*_j}(p')+E_{\pi^c}
(\vep+\vep')-M\Bigr),&& i\ne j.
\end{array}
\right.
\label{D2}
\ee

For convenience,  the energy $E$ is counted relative to the neutral two-body threshold,
\be
M=m_{*0}+m_0+E,
\ee
while the energies of the individual particles are
\be
E_{\pi_i}(\vep)=\sqrt{\vep^2+m_{\pi_i}^2},\quad E_{D_i}(\vep)=\sqrt{\vep^2+m_i^2},\quad
E_{D^*_i}(\vep)=\sqrt{\vep^2+m_{*i}^2}.
\ee

The OPE potential (\ref{Vmn}) interrelates the four $D$-meson channels defined as
\be
|0\rangle=D^0\bar{D}^{*0},\quad|\bar{0}\rangle=\bar{D}^0 D^{*0},\quad
|c\rangle=D^+ D^{*-},\quad|\bar{c}\rangle=D^- D^{*+}.
\ee

Then the system of coupled-channel Lippmann-Schwinger equations for the $D\bar{D}^*$ $t$-matrix elements
$a_{00}^{nn'}(\vep,\vep')$ and $a_{c0}^{nn'}(\vep,\vep')$ in the $C$-even channel has the form
\cite{Baru:2013rta}
\bea
\left\{
\begin{array}{l}
\hspace*{-2pt}a_{00}^{nn'}(\vep,\vep')=\lambda_0 V_{00}^{nn'}(\vep,\vep')
-\ds\sum_{i=0,c}\lambda_i\int d^3k V_{0i}^{nm}(\vep,\vek)\frac{1}{\Delta_i(\vek)}a_{i0}^{mn'}(\vek,\vep'),\\
\hspace*{-2pt}a_{c0}^{nn'}(\vep,\vep')=\lambda_c V_{c0}^{nn'}(\vep,\vep')
-\ds \ds\sum_{i=0,c}\lambda_i\int d^3k V_{ci}^{nm}(\vep,\vek)\frac{1}{\Delta_i(\vek)}a_{i0}^{mn'}(\vek,\vep'),
\end{array}
\right.
\label{aa}
\eea
where $\lambda_0=\langle 0|\vec\tau_1\cdot\vec\tau_2|\bar{0}\rangle=\langle
c|\vec\tau_1\cdot\vec\tau_2|\bar{c}\rangle=1$ and $\lambda_c=\langle
0|\vec\tau_1\cdot\vec\tau_2|\bar{c}\rangle=\langle
c|\vec\tau_1\cdot\vec\tau_2|\bar{0}\rangle=2$ are the isospin factors for the $\pi^0$- and $\pi^\pm$-exchange,
respectively.

The partial wave projections of potential (\ref{Vmn}) on the relevant $^3S_1$ and $^3D_1$ partial waves read
($x=\cos\theta$ where $\theta$ is the angle between the momenta $\vep$ and $\vep'$)
\begin{eqnarray*}
V_{ij}^{SS}(p,p')&=&C_0+\frac{1}{6}\int^1_{-1}F_{ij}(p,p',x)\left( p^2+ p'^2+
2 pp'x\right)dx,\\
V_{ij}^{SD}(p,p')&=&-\frac{\sqrt{2}}{6}\int^1_{-1}F_{ij}(p,p',x)
\left[p'^2+ p^2\left(\frac32x^2-\frac12\right)+2 pp'x\right]dx,\\
V_{ij}^{DS}(p,p')&=&-\frac{\sqrt{2}}{6}\int^1_{-1}F_{ij}(p,p',x)\left[p^2+
p'^2\left(\frac32x^2-\frac12\right)+2pp'x\right]dx,\\
V_{ij}^{DD}(p,p')&=&\frac{1}{3}\int^1_{-1}F_{ij}(p,p',x)\left[(p^2+p'^2)\left(\frac32
x^2-\frac12\right)+\frac{11}{10}pp'x
+\frac{9}{10}pp'\left(\frac52 x^3-\frac32 x\right)\right]dx.
\end{eqnarray*}

 Because of the $P$-wave nature of the $D^*\to D\pi$ vertex, the $D\pi$ loop operator
$\Sigma(s,m_*,m,m_\pi)$ diverges and it is subject to renormalisation. The necessary details of the
renormalisation procedure are given in
Appendix~\ref{app:renorm} while here we only quote the final result for the inverse  two-body  propagators $\Delta_0$ and $\Delta_c$ entering system of equations (\ref{aa}):
\bea\nonumber
\ds\Delta_0(\vep)&=&\frac{E_{D^0}(\vep)E_{D^{*0}}(\vep)}{m_0m_{*0}}\left[\frac{E_{D^0}(\vep)+E_{D^{*0}}(\vep)-M}{\zeta}
\right.\nonumber\\
&-&\left.\frac{\tilde{\Sigma}_R(s, m_{*0},m_{\pi^0},m_0)+2\tilde{\Sigma}_R(s, m_{*0},m_{\pi_ c},m_c)+i
m_{*0}\varGamma(D^{*0}\to D^0\gamma)}{2E_{D^{*0}}(\vep)}\right],\nonumber\\[-2mm]
\label{DD*prop2}\\[-2mm]
\ds\Delta_c(\vep)&=&\frac{E_{D^c}(\vep)E_{D^{*c}}(\vep)}{m_cm_{*c}}
\left[\frac{E_{D^c}(\vep)+E_{D^{*c}}(\vep)-M}{\zeta}\right.\nonumber\\
&-&\left.
\frac{\tilde{\Sigma}_R(s, m_{*c},m_{\pi^0},m_c)+2\tilde{\Sigma}_R(s, m_{*c},m_{\pi_ c},m_0)}{2E_{D^{*c}}(\vep)}\right],
\nonumber
\eea
where
\be
s=m_*^2+2E_{D^*}(\vep)(M- E_D(\vep)-E_{D^*}(\vep))
\label{runs}
\ee
for the off-shell $D^*$ resonance and $\tilde{\Sigma}_R(s,m_*,m,m_\pi)$ is the renormalised loop operator defined at the
``running'' pion mass, that, in particular, brings about the quantity $\zeta$,
\be
\zeta^{-1}\equiv 1-\frac{g_R^2}{384\pi^2}\ln\frac{m_\pi^2}{\oo{m_\pi}},
\ee
with the renormalised coupling constant defined as (see Appendix~\ref{app:renorm})
\be
g_R=g_c\frac{\sqrt{m_0m_{*0}}}{f_\pi}.
\label{ggcgam}
\ee

For future discussion of the static approximation, we also consider a simplified case of the constant width which corresponds to the substitution
of the constant $s=m_*^2$ in Eq.~(\ref{DD*prop2}) instead of the ``running'' $s$, as given in Eq.~(\ref{runs}).

We are now in a position to introduce the power counting in the parameter $\xi=m_\pi/m_\pi\0$.
The   $m_\pi$-dependence of the coupling constant $g_c$ is extracted from
Ref.~\cite{Becirevic:2012pf} and is discussed in Ref.~\cite{Baru:2013rta}.
In particular, at  LO $g_c$ remains constant while at NLO it acquires corrections of the order of
$m_\pi^2$. Similarly, $f_\pi$, the masses of the $D$ and $D^*$ mesons, and the decay width $\varGamma(D^{*0}\to
D^0\gamma)$ take their respective physical values in the LO calculation. The central issue of this work is related to
the $m_\pi$-dependence of the contact interaction $C_0$ --- see Eq.~(\ref{Vmn}). Since the nature of this interaction is
obscure, the dependence $C_0(m_\pi)$ can only be guessed using the principle of naturalness. Below,
we discuss in detail generalisation of the corresponding approach developed in Ref.~\cite{Baru:2013rta}. Meanwhile,
regardless of the particular source of the dependence $C_0(m_\pi)$ it
only appears at NLO, so that the contact interaction remains constant at LO and, therefore, the problem is fully
fixed to provide a prediction for the behaviour of the $X$-pole as the pion mass leaves the physical point.
Furthermore, the pion-mass dependence at LO occurs only due to the pion energies in the $D\bar D \pi$ propagator and
pion-mass effects in the renormalised loop --- both are the equally important parts of the three-body $D\bar D \pi$
dynamics.
To finalise the setup of the problem, we quote the values of various parameters at the physical pion mass used in the
calculations. In particular, in the physical limit of $m_\pi=m_\pi\0$ one has  $f_\pi\0=92.4$ MeV, then
\begin{eqnarray*}
m_{\pi^0}\0=134.98~\mbox{MeV},\quad m_0\0=1864.84~\mbox{MeV},\quad m_{*0}\0=2006.97~\mbox{MeV},\\
m_{\pi^c}\0=139.57~\mbox{MeV},\quad m_c\0=1869.62~\mbox{MeV},\quad m_{*c}\0=2010.27~\mbox{MeV},
\end{eqnarray*}
and the values
\be
\varGamma\0(D^{*0}\to D^0\pi^0)=42~\mbox{keV},\quad\varGamma\0(D^{*0}\to D^0\gamma)=21~\mbox{keV}
\ee
can be deduced from the data for the charged $D^*$ decay modes \cite{PDG}.
The physical values of the couplings introduced above are \cite{Baru:2013rta}
\be
g_c^{\rm ph}=0.61,\quad g_R^{\rm ph}=12.7.
\ee

\begin{figure}[t]
\centerline{
\epsfig{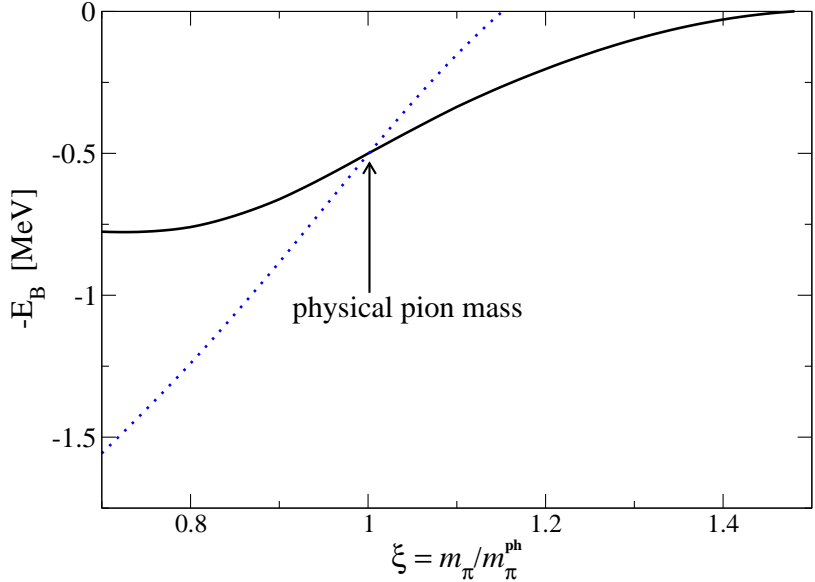}}
\caption{Pion mass dependence of the $X(3872)$ binding energy at LO. The results of the full dynamical theory with
3-body effects included (black solid curve)
are confronted with the simplified formulation with static OPE (blue dotted curve). }
\label{figLO}
\end{figure}

\section{Results and Discussions}\label{results}

We are now in a position to discuss the results for the pion mass (or, equivalently, light-quark mass) dependence of the
$X(3872)$ binding energy $E_B(m_{\pi})$.
We start from the discussion of the LO results. As was explained above, the contact term $C_0$ is $m_\pi$-independent at this order, so once it is
adjusted to reproduce the binding energy at the physical pion mass (for definiteness we set $E_B(m_\pi\0)=0.5$~MeV), the
scattering amplitude can be calculated for unphysical pion masses without loss of renormalisability of the LO equations.
Therefore, at  LO of our EFT, the dependence $E_B(m_{\pi})$ can be predicted in a parameter-free way --- see
Fig.~\ref{figLO}.
At this order, the pion mass dependence of $E_B$ originates only from the pionic effects in the OPE potential and from
those in the
renormalised selfenergy loops $\tilde\Sigma_R$ --- see Eq.~(\ref{DD*prop2}). The binding energy at LO
demonstrates a clear tendency to decrease with the $m_\pi$ growth. Note that a similar behaviour of the binding energy
was observed in Ref.~\cite{Epelbaum:2013ij} for the deuteron. Furthermore, the slope of the binding
energy in $m_\pi$ at the physical point, $(\partial E_B/\partial m_\pi){\big |}_{m_\pi=m_\pi^{\rm ph}}$, exhibits a
strong sensitivity to the 3-body $D\bar{D}\pi$ effects. In particular, neglecting the
3-body dynamics (the so-called static OPE) results in a much steeper fall of the binding energy --- compare the dotted
(blue) line
versus the solid (black) in Fig.~\ref{figLO}.

\begin{figure}[t]
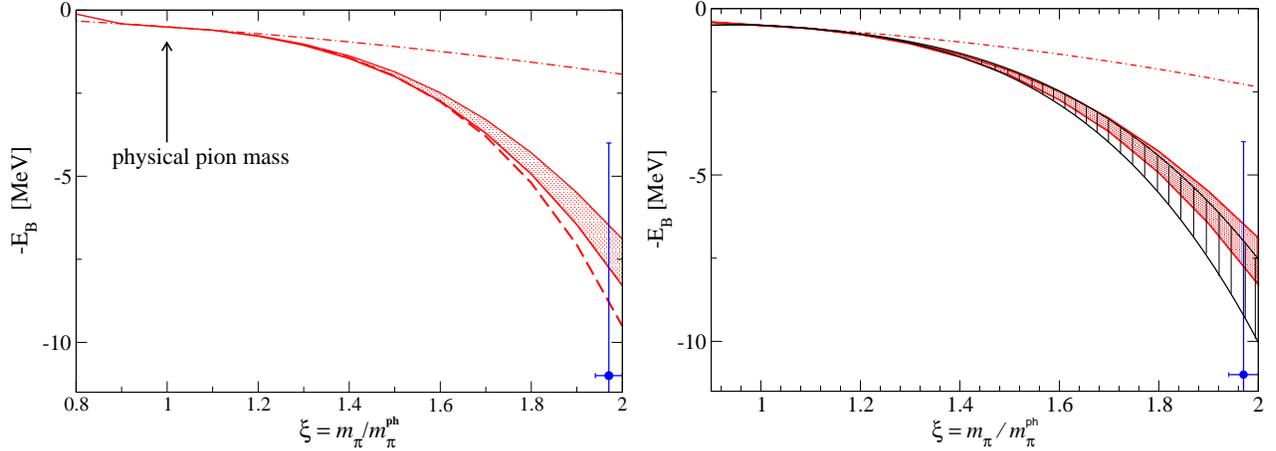

\centerline{
\hspace*{0mm}\epsfig{file=NLO_dynam_vs_pionless_paper.eps,width=0.5\textwidth}\hspace*{1mm}
\epsfig{file=NLO_dynam_vs_PLB_finiteLam_paper.eps,width=0.5\textwidth}}
\caption{Pion mass dependence of the $X(3872)$ binding energy at  NLO. Left panel: the dashed line is for the
perturbative treatment of the $m_\pi$-dependent contact operator at  NLO while the red dotted band represents the
nonperturbative results employing resonance saturation. Right panel: comparison of the results obtained
in the heavy-meson formulation of Ref.~\cite{Baru:2013rta} with the finite cut-off $\Lambda\in$ [500 MeV, 700 MeV]
(black hatched band) with the nonperturbative results of the current study employing resonance saturation and the
cut-off $\Lambda\to\infty$ (red dotted band). The dashed-dotted line in both panels corresponds to the calculation
without pions. The (blue) dot with the error bars shows the result of the lattice calculation of
Ref.~\cite{Padmanath:2015era}.}\label{figNLO}
\end{figure}

Since no real experiment is possible for unphysical pion masses, the only source of information on the $X$ pole fate for
the $m_\pi$'s exceeding the physical pion mass is provided by lattice simulations. Such calculations are indeed being
performed
and most of them predict an increase of the binding energy with the $m_\pi$ growth. For example, different lattice
collaborations observe this type of behaviour for the deuteron --- see, for example,
Refs.~\cite{Beane:2011iw,Beane:2012vq,Yamazaki:2012hi} and references therein\footnote{
On the other hand, the HAL QCD Collaboration found no bound state in the $NN$ $^3S_1$-$^3D_1$ channel
\cite{Inoue:2011ai}.}. Also the first lattice calculations for the $X(3872)$ indicate the existence of a stronger bound
$X$ for $m_\pi>m_\pi^{\rm ph}$ \cite{Prelovsek:2013cra,Lee:2014uta,Padmanath:2015era}. Although these results still
suffer from potentially large finite-range corrections, as pointed out in
Ref.~\cite{Jansen:2015lha} within X-EFT, they raise an important question of whether such a behaviour of the binding
energy can be understood theoretically. To this end, we go beyond  LO and proceed to  NLO thus including
corrections quadratic in $m_\pi$. In particular, we allow for an $m_\pi$-dependence of the short-range interaction
which therefore goes away from its physical value. Thus, we consider (for simplicity all indices are omitted)
\be
V_{\rm NLO}=V_{\rm OPE}(\vep,\vep',\xi)+C_0+D(\xi^2-1),\quad \xi=m_\pi/m_\pi^{\rm ph},
\label{Dxi2}
\ee
where the first two terms on the right-hand-side stand for the LO potential (\ref{Vmn}) while the last term accounts for
our ignorance of other
dynamical scales but those related to the OPE. As was discussed in the previous Section,
renormalisability of the theory requires all operators beyond the LO to be included perturbatively.
Following Ref.~\cite{Baru:2013rta}, we fix the unknown coefficient $D$ to the slope of the binding energy at the physical
pion mass which is therefore considered as an additional input quantity.
For example, in Fig.~\ref{figNLO} we illustrate the behaviour of the binding energy for the slope
$(\partial E_B/\partial m_\pi^2){\big |}_{m_\pi=m_\pi\0} \approx \, E_B^{\rm ph}/{m_\pi^{\rm ph}}^2$ --- see the dashed
curve in the left panel.
While the sign of the slope was fixed to provide a growth of the binding energy with the pion mass, its magnitude was
chosen
to comply with naturalness. Specifically, we assume that the shift of the binding
energy $\delta E_B\sim E_B^{\rm ph}$ for $\delta m_\pi\sim m_\pi^{\rm ph}$ can be interpreted as natural. Indeed, the
slope predicted at  LO due to OPE fulfils this criterion:
$(\partial E_B/\partial m_\pi^2){\big |}_{m_\pi=m_\pi\0} \approx - 1.5 \, E_B^{\rm ph}/{m_\pi^{\rm ph}}^2$. Therefore,
to study the case of a stronger bound $X$, we fix the slope to be
$(\partial E_B/\partial m_\pi^2){\big |}_{m_\pi=m_\pi\0} \approx \, E_B^{\rm ph}/{m_\pi^{\rm ph}}^2$.
Interestingly, in a theory with the same polynomial behaviour of the contact operator but without pions, one would
observe a
much flatter behaviour $E_B(m_\pi)$ for the same slope $(\partial E_B/\partial
m_\pi^2){\big |}_{m_\pi=m_\pi^{\rm ph}}$, as shown by the dashed-dotted curve.
The difference between the two curves demonstrates the role of dynamical pions as an explicit long-range degree of
freedom.
As seen from Fig.~\ref{figNLO}, the contact interaction provides a smooth background for a rapidly varying pion-mass
dependence stemming from OPE.
Therefore, integrating out pions and the corresponding 3-body soft scales while still trying to,
at least partially, compensate for neglecting these long-range effects, one would inevitably arrive at unnaturally large
$m_\pi$-dependent coefficients accompanying short-range operators.

On the other hand, one may question a justification of the perturbative inclusion of the $m_\pi$-dependent short-range
interaction in Eq.~(\ref{Dxi2}).
Given the shallowness of the physical $X$ state, even a small variation of the slope within its natural range, as
discussed above, has a sizeable impact on the $m_{\pi}$-dependence of the binding energy.
In order to verify the validity of the perturbative approach, we employ
resonance saturation to model higher-order contact interactions by means of a heavy-meson exchange. In
particular, we consider the NLO potential in the form
\be
V_{\rm NLO}=V_{\rm OPE} (\vep,\vep',\xi)+C_0 +\beta\frac{g_c^2}{(4\pi f_\pi)^2}\frac{({\bm \varepsilon}\cdot\veq)
({\bm\varepsilon}^*\cdot\veq)}{\veq^2+M^2+\delta M^2(\xi^2-1)},
\label{satur}
\ee
where $\veq=\vep+\vep'$ and the scale $M$ is varied in the range $M=600...800$~MeV that corresponds to a typical
heavy-meson mass. The  parameter $\beta$ accounts  for   the  difference  in the strength of  the  heavy-meson exchange potential relative to that    of   OPE.  
It  is expected to take values around unity and it could be, in principle,
adjusted to the $D\bar{D}^*$ effective range. However, given that the latter is unknown, $\beta$ is varied within a
suitable range of values from 1 to 2 which we treat as natural. The term $\delta M^2(\xi^2-1)$ in the denominator
accounts for the pion mass dependence of the heavy meson mass\footnote{Notice,   a similar EFT approach based on the resonance saturation 
hypothesis was  used in Ref.~\cite{Berengut:2013nh} to constrain the pion mass dependence of the short-range  $NN$ forces.} 
with $\delta M$ adjusted to the slope
$(\partial E_B/\partial m_\pi^2){\big |}_{m_\pi=m_\pi^{\rm ph}}$. We have verified that for $\delta M=0$,
the dependence $E_B(m_\pi)$ is basically indistinguishable from the LO one that
confirms the results to be insensitive to the details of the short-range interaction, as expected.
The form of the NLO potential (\ref{satur}) ensures that the corresponding scattering amplitude is renormalisable so
that the NLO calculations can be carried out in the same way as the LO ones. Then, fixing the slope as before,
$(\partial E_B/\partial m_\pi^2){\big |}_{m_\pi=m_\pi\0} \approx \, E_B^{\rm ph}/{m_\pi^{\rm ph}}^2$, we obtain the
dotted (red) band in Fig.~\ref{figNLO}.
A very good agreement between this band and the dashed curve in the considered range of pion masses
confirms that higher-order effects originating from the non-perturbative resummation of  pion-mass-dependent
short-range interactions
are minor.
Therefore, the perturbative treatment of the pion-mass-dependent short-range interaction, as given
by Eq.~(\ref{Dxi2}), is indeed justified even though it brings about a new effect --- the slope of the binding energy
may change its sign compared
to the LO result depicted in Fig.~\ref{figLO}.

It is also instructive to compare the results of the present study with those obtained in Ref.~\cite{Baru:2013rta}
in the heavy-meson formulation with a finite cut-off --- see the right panel in Fig.~\ref{figNLO}.
At NLO, both approaches are consistent with each other as may be expected since both
formulations are justified in general. Meanwhile, as was already discussed above, the LO
equation is explicitly renormalisable and predictive in the present formulation in contrast to the approach of
Ref.~\cite{Baru:2013rta}, where the
$m_\pi$-dependent contact interaction requires an additional input to be included to maintain the renormalisability
of the scattering amplitude. In addition, the requirements of naturalness are much easier to formulate and to apply in
the
current approach since the dependence of the results on the cut-off is eliminated.
Thus, we emphasise that the role of pion dynamics can be understood in a much more
transparent way using the explicitly renormalisable theory which is free of finite cut-off artefacts.

To further clarify the role of the dynamical pions at  NLO, we
assume that there exist gedanken lattice data at unphysical pion masses. These data could be used to adjust the
parameters of the short-range potential. Then, once the short-range physics is fixed, the theory can be extrapolated
to the physical point in $m_\pi$ and confronted with the experimental data. In particular, if the
lattice calculations provide two measurements of the binding energy of the $X$ made for two unphysically large pion
masses then the suggested approach allows us to establish the correct extrapolating formula to the physical point and
thus to predict the corresponding value $E_B(m_\pi\0)$. In addition, information on the behaviour of the short-range
interactions in the $X$,  which can, in this way, be extracted from the lattice data,  may shed light on the nature of the
binding mechanisms in the $X$. This establishes an important link between the EFT approach and
lattice simulations for hadronic molecule states.

\begin{figure}[t]
\centerline{\epsfig{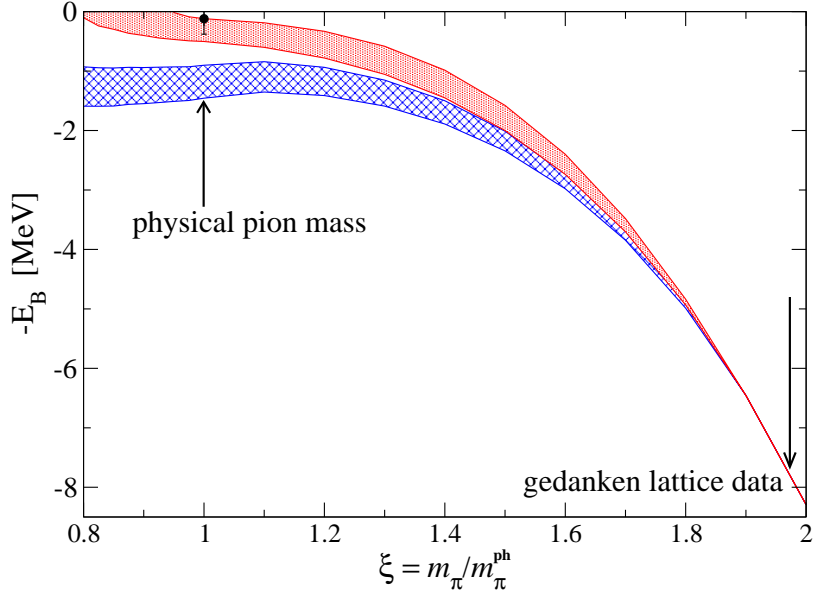}}
\caption{Pion mass dependence of the $X(3872)$ binding energy. The red dotted band is for the full calculation
with dynamical pions at NLO while the blue crossed band is for the static OPE. }\label{figextr}
\end{figure}

As an illustration, the chiral extrapolations for the two theories, the one with
dynamical pions and the one with the static OPE, are compared with each other in Fig.~\ref{figextr}. 
For definiteness the gedanken lattice result is taken at $m_\pi=2m_\pi^{\rm ph}$, as indicated by the arrow in
Fig.~\ref{figextr}. If the slope is
chosen such that the theory with dynamical pions
provides the correct extrapolation to the physical pion mass, the static theory with the same slope yields a significant
overbinding of the $X(3872)$, by more than a factor of 3 in the binding energy. In addition, one can see from
Fig.~\ref{figextr} that the extrapolation curve from an unphysically large pion mass, close to the
values used on the lattices, to the physical point is nontrivial and the corresponding extrapolating behaviour
has a strong curvature. This illustration emphasises the importance of the 3-body effects for the chiral extrapolations
for the $X$ and puts in question possibility of using any simple ansatz for the extrapolation formula.

\section{Summary and conclusions}\label{summary}

In this work we developed an explicitly renormalisable framework
to study chiral extrapolations of the binding energy of the $X(3872)$ beyond the physical pion mass.
This approach is free of the finite cut-off artefacts  which is a precondition  for a systematic control over the pion-mass dependence from the 
short-range interactions. The pertinent results of our work can be summarised
as follows.  First,  the interplay between the long- and short-range forces in
the $X$ appears to be quite nontrivial,  as was already pointed out in Ref.~\cite{Baru:2013rta}. 
If the $X$ turns out to be less bound for the pion masses exceeding
its physical value, the $m_\pi$-dependence of the $X$ binding energy is entirely governed by the explicit pion-mass
dependence of the OPE potential. On the other hand, a stronger bound $X$ would signal the importance of the
$m_{\pi}$-dependent short-range interactions in addition to pionic effects. Confronting our results with those
of the lattice simulations could allow one to extract valuable information on such short-range interactions and,
possibly, to  disclose  
the nature of the binding forces in the $X$.

Secondly, our findings are of a practical value for the lattice simulations since they open the way
 to override the gap between the unphysically large pion masses used on the lattices and the physical limit.
It follows from our results that the corresponding interpolating curve has a strong curvature and it is strongly
affected by the 3-body effects in the $X$.

Last but not least, the approach developed in this paper can also be adapted to other near-threshold states,
the $X(3872)$ being just the most prominent and therefore the most extensively studied one.

\begin{acknowledgments}

We would like to thank  Christoph Hanhart  and Ulf-G. Mei\ss ner  for a  careful  reading of the manuscript  and  valuable remarks. 
This work is supported by   the ERC project 259218 NUCLEAREFT, the DFG and the NSFC through funds
provided to the Sino-German CRC 110 ``Symmetries
and the Emergence of Structure in QCD'', the Russian Science Foundation (Grant No. 15-12-30014)
 and by the Georgian Shota Rustaveli National Science Foundation (grant FR/417/6-100/14).

 \end{acknowledgments}

\appendix

\section{The $D\pi$ loop operator and the $D^*$ propagator}\label{app:renorm}

Consider an unstable vector mesonic state which decays in the $P$ wave into a pair of (pseudo)scalar mesons. We start
from its inverse propagator
\be
D(s)=s-m_0^2+\Sigma(s),
\ee
where $s$ is the invariant energy ($s=p^2$), $m_0$ stands for the ''bare`` mass, and $\Sigma(s)$
denotes the selfenergy loop operator. The one-loop contribution to the selfenergy has
the form\footnote{ Vector meson self-energies in chiral EFT were first discussed in Ref.~\cite{Bruns:2004tj}.}
\be
\Sigma(s)=-g_0^2 I_{\mu \nu}\, \varepsilon^{\mu} {\varepsilon^*}^\nu,
\label{RSE}
\ee
where $g_0$ is a dimensionless bare coupling constant and $\varepsilon$'s stand for the polarisation vectors of the
unstable meson such that $(\varepsilon \cdot {\varepsilon^*})=-1$ and $(\varepsilon \cdot p)=0$.

The loop integral corresponding to the decay of the vector meson into two mesons of the masses $m_1$ and $m_2$ reads
\be
I_{\mu\nu}= i\int\frac{d^nk}{(2\pi)^n}\frac{k_\mu k_\nu}{(k^2-m_1^2)((k+p)^2-m_2^2)}=Ag_{\mu\nu}+B\frac{p_\mu p_\nu}{p^2}
\label{Imn}
\ee
and, due to the property $(\varepsilon \cdot p)=0$, only the function $A$ is relevant for the loop operator (\ref{RSE}).
Since the loop integral (\ref{Imn}) diverges quadratically in the limit of $n\to 4$, the coefficients
$A_0$ and $A_1$ in the Taylor series of the function $A(s)$,
\be
A(s)=A_0+A_1s+A_{\rm reg}(s),
\label{As}
\ee
 are singular in this limit, while the residual function $A_{\rm reg}(s)$ is regular.
By a straightforward calculation in the dimensional regularisation scheme, it is easy to find that
\bea
&\ds A_0=\frac{1}{192\pi^2}\left((2m_2^2+m_1^2)C(m_2)+(2m_1^2+m_2^2)C(m_1)\right),\quad
A_1=-\frac{1}{384\pi^2}\left(C(m_1)+C(m_2)\right),&\nonumber\\[-2mm]
\label{A1sing}\\[-2mm]
&\ds A_{\rm reg}(s)=\frac{1}{12s}\left[\frac{m_1^4-m_2^4}{16\pi^2}\ln\frac{m_1}{m_2}+\lambda(s,m_1^2,m_2^2)I_2^{\rm
reg}(m_1,m_2,s)\right],&\nonumber
\eea
where
\beas
I_2^{\rm reg}(s,m_1,m_2)&=&-\frac{1}{16\pi^2}\left(-1+\frac{m_1^2-m_2^2}{s}\ln\frac{m_1}{m_2}\right.\\[2mm]
&+&\left.\frac{\lambda^{1/2}(s,m_1^2,m_2^2)}{s}\ln\frac{m_1^2+m_2^2-s-\lambda^{1/2}(s,m_1^2,m_2^2)}{2m_1m_2}
\right)
\eeas
and
$$
C(m)=\left(\frac{1}{\epsilon}-\log(4\pi)+\gamma_E-1\right)+\ln\frac{m^2}{\mu^2},\quad \epsilon=\frac12(4-n)\to 0,\quad
\gamma_E=-\Gamma'(1)\approx 0.5772,
$$
while $\mu$ is the scale in dimensional regularisation and  the triangle function is defined in the standard way, as
\be\label{lambda}
\lambda(x,y,z)=x^2+y^2+z^2-2xy-2xz-2yz.
\ee

Then, the one-loop contribution to the selfenergy in Eq.~(\ref{RSE}) takes the form
\be
\Sigma(s)=g_0^2A(s)=g_0^2(A_0+A_1s+A_{\rm reg}(s)),
\ee
and it is subject to renormalisation which we perform by expanding $\Sigma(s)$ near the renormalised vector meson
mass $m_R$,\footnote{In general it is preferable to relate the renormalised mass to the complex pole of the propagator
\cite{Willenbrock:1990et,Willenbrock:1991hu,Sirlin:1991fd,Gegelia:1992kj,Gambino:1999ai,Djukanovic:2007bw}. However, at
the one-loop level it is sufficient to use the real part of the inverse propagator since the difference between the two
approaches occurs only starting from the two-loop order.}
\be
\Sigma(s)={\rm Re}\Sigma(m_R^2)+{\rm Re}\Sigma'(m_R^2)(s-m_R^2)+\Sigma_{\rm reg}(s),
\ee
where
\be
{\rm Re}\Sigma(m_R^2)=g_0^2(A_0+A_1m_R^2+{\rm Re}A_{\rm reg}(m_R^2)),\quad
{\rm Re}\Sigma'(m_R^2)=g_0^2(A_1+{\rm Re}A_{\rm reg}'(m_R^2)),
\ee
while
\be
\Sigma_{\rm reg}(s)=\Sigma(s)-{\rm Re}\Sigma(m_R^2)-{\rm Re}\Sigma'(m_R^2)(s-m_R^2)
\equiv g_0^2A_R(s),
\label{SR}
\ee
with
\be
A_R(s)=A_{\rm reg}(s)-{\rm Re}A_{\rm reg}(m_R^2)-{\rm Re}A_{\rm reg}'(m_R^2)(s-m_R^2).
\ee
Notice that $A_R(s)$ is finite and it does not depend on the auxiliary regularisation scale $\mu$.

Defining the renormalised mass such that $m_R^2=m_0^2-{\rm Re}\Sigma(m_R^2)$, we have:
\beas
D(s)=(s-m_0^2+{\rm Re}\Sigma(m_R^2))+{\rm Re}\Sigma'(m_R^2)(s-m_R^2)+\Sigma_R(s)=Z^{-1}(s-m_R^2)+g_0^2A_R(s),
\eeas
with
\be
Z^{-1}\equiv 1+{\rm Re}\Sigma'(m_R^2)=1+g_0^2(A_1+\mbox{Re}A'_{\rm reg}(m_R^2)),
\label{Zdef}
\ee
where it was used that $A_1$ is real.

Consider now the combination entering the system of equations for the scattering amplitudes (\ref{aa}),
\be
\frac{g_0^2}{D(s)}=\frac{g_0^2}{Z^{-1}(s-m_R^2)+g_0^2 A_R(s)}=\frac{g_R^2}{s-m_R^2+\Sigma_R(s)},\quad
\Sigma_R(s)=g_R^2 A_R(s),
\label{form10}
\ee
where the renormalised coupling constant $g_R$ is defined as
\be
g_R^2\equiv Z g_0^2.
\label{gamren}
\ee

In particular, for the case of the $D^{*0}$, an obvious identification of the parameters is $m_R=m_{*0}$, $m_1=m_0$,
$m_2=m_{\pi^0}$. Then, using Eq.~(\ref{D*width2}) and the standard relation between the loop operator and the width,
\bea
\varGamma(D^{*0}\to D^0\pi^0)&=&\frac{1}{m_{*0}}\mbox{Im}{\Sigma}_R(s=m_{*0}^2,
m_R=m_{*0},m_{\pi^0},m_0)\nonumber\\[-2mm]
\\[-2mm]
&=&\frac{g_R^2}{m_{*0}}\mbox{Im}A_{\rm reg}(s=m_{*0}^2, m_R=m_{*0},m_{\pi^0},m_0),\nonumber
\eea
where $\mbox{Im}A_{\rm reg}$ can be found from Eq.~(\ref{A1sing}) to be
\be
\mbox{Im}A_{\rm reg}(s=m_{*0}^2, m_R=m_{*0},m_{\pi^0},m_0)=\frac{q^3}{24\pi m_{*0}},\quad
q=\frac{1}{2m_{*0}}\lambda^{1/2}(m_{*0}^2,m_0^2,m_{\pi^0}^2),
\ee
one arrives at the following relation between the couplings $g_R$ and $g_c$:
\be
g_R=g_c\frac{\sqrt{m_0m_{*0}}}{f_\pi},
\label{ggcgam0}
\ee
that completes renormalisation programme at the physical point.

Away from the physical value of the pion mass one can write:
\bea
D(s)&=&s-m_0^2+\Sigma(s)=s-m_R^2+\Bigl[\Sigma(s)-{\rm Re}\Sigma(m_R^2)\Bigr]
=s-m_R^2+g_0^2\Bigl[A(s)-{\rm Re}A(m_R^2)\Bigr]\nonumber\\
&=&s-m_R^2+g_0^2\Bigl[A_1(s-m_R^2)+A_{\rm reg}(s)-{\rm Re}A_{\rm reg}(m_R^2)\Bigr]\\
&=&s-m_R^2+g_0^2\Bigl[A_1(s-m_R^2)+{\rm Re}A_{\rm reg}^{\prime\rm
ph}(m_R^2)(s-m_R^2)+\tilde{A}_R(s)\Bigr],\nonumber
\eea
where
\be
\tilde{A}_R(s)=A_{\rm reg}(s)-{\rm Re}A_{\rm reg}(m_R^2)-{\rm Re}A_{\rm reg}^{\prime \rm ph}(m_R^2)(s-m_R^2),
\ee
that is, $\tilde{A}_R(s)$ is defined with the derivative in the subtracted term evaluated at the physical point. This
allows one to avoid $m_\pi$-dependence in the renormalisation factor $Z$ and to preserve its definition in the form of
Eq.~(\ref{Zdef}). Then
\bea
D(s)&=&(s-m_R^2)\Bigl(1+g_0^2(A_1+{\rm Re}A_{\rm reg}^{\prime\rm
ph}(m_R^2))\Bigr)+g_0^2\tilde{A}_R(s)\nonumber\\
&=&(s-m_R^2)\Bigl(1+g_0^2(A_1-A_1^{\rm ph})+g_0^2(A_1^{\rm ph}+{\rm Re}A_{\rm reg}^{\prime\rm
ph}(m_R^2))\Bigr)+g_0^2\tilde{A}_R(s)\nonumber\\
&=&(s-m_R^2)\Bigl(g_0^2(A_1-A_1^{\rm ph})+Z^{-1}\Bigr)+g_0^2\tilde{A}_R(s)\\
&=&Z^{-1}\Bigl[(s-m_R^2)\Bigl(1+g_R^2(A_1-A_1^{\rm ph})\Bigr)+g_R^2\tilde{A}_R(s)\Bigr]\nonumber\\
&=&Z^{-1}\left[(s-m_R^2)\left(1-\frac{g_R^2}{384\pi^2}\ln\frac{m^2}{\oo{m}}\right)+g_R^2\tilde{A}_R(s)\right],
\nonumber
\eea
where the definition of the renormalised coupling $g_R$, Eq.~(\ref{gamren}), and the explicit form of $A_1$,
Eq.~(\ref{A1sing}), were used. Therefore for the ``running'' pion mass instead of Eq.~(\ref{form10}) one has
\be
\frac{g_0^2}{D(s)}=\frac{g_0^2}
{Z^{-1}\left[(s-m_R^2)\zeta^{-1}+g_R^2\tilde{A}_R(s)\right]}=
\frac{g_R^2}{(s-m_R^2)\zeta^{-1}+\tilde{\Sigma}_R(s)}\equiv\frac{g_R^2}{D_R(s)},
\label{form2}
\ee
where
\be
\zeta^{-1}\equiv 1-\frac{g_R^2}{384\pi^2}\ln\frac{m_\pi^2}{\oo{m_\pi}}
\ee
and
\be
\tilde{\Sigma}_R(s)=g_R^2\left[A_{\rm reg}(s)-{\rm Re}A_{\rm reg}(m_R^2)-{\rm Re}A_{\rm reg}^{\prime \rm
ph}(m_R^2)(s-m_R^2)\right].
\label{SigR}
\ee

After integrating over the zeroth component in the loop   (one picks the $D$-meson to be on energy shell),  one  finds   
\bea
s-m_{*0}^2\approx 2E_{D^{*0}}(\vep)(p_0-E_{D^{*0}}(\vep))=
2E_{D^{*0}}(\vep)(M- E_{D^0}(\vep)-E_{D^{*0}}(\vep)).
\eea

Therefore, one arrives at the formula
\be
D_{D^{*0}}(\vep)=\frac{2E_{D^{*0}}(\vep)}{\zeta}\Bigl[M-E_{D^0}(\vep)-E_{D^{*0}}(\vep)\Bigr]+\tilde{\Sigma}_R(\vep),
\ee
for the renormalised inverse $D^{*0}$ propagator valid for unphysical pion masses.   Obviously, a similar formula holds for the inverse $D^{*c}$ propagator for the 
charged particles, as well.

The inverse  two-body  propagators $\Delta_0$ and $\Delta_c$ entering system of equations (\ref{aa}) can be written
as
\begin{equation}
\Delta_0(\vep)=-\frac{E_{D^0}(\vep)}{m_0} \, \frac{ D_{D^{*0}} (\vep)}{2m_{*0}},\quad
\Delta_c(\vep)=-\frac{E_{D^c}(\vep)}{m_c} \, \frac{ D_{D^{*c}} (\vep)}{2m_{*c}}.
\label{RPr}
\end{equation}

\end{document}